\documentclass[prb,12pt]{revtex4}
\usepackage[english]{babel}
\usepackage[intlimits]{amsmath}
\usepackage[dvips]{graphicx}
\usepackage{epsfig}

\begin{document}

\begin{titlepage}

\title{Translocation of polymers with folded configurations across nanopores}
\author{Stanislav Kotsev and Anatoly B. Kolomeisky}
\affiliation{Department of Chemistry, Rice University, Houston, TX 77005-1892}

\begin{abstract}
The  transport of polymers with folded configurations across membrane pores is investigated theoretically by analyzing simple discrete stochastic models. The translocation dynamics is viewed as a sequence of two events: motion of the folded segment through the channel followed by the linear part of the polymer. The transition rates vary for the folded and linear segments because of different interactions between the polymer molecule and the pore. It is shown that the translocation time depends non-monotonously on the length of the folded segment for short polymers and weak external fields, while it becomes monotonous for  long molecules and large fields. Also, there is a critical interaction between the polymers and the pore that separates two dynamic regimes. For stronger interactions the folded polymer moves slower, while for weaker interactions the linear chain translocation is the fastest. In addition, our calculations show that the folding does not change the translocation scaling properties of the polymer. These phenomena can be explained by the interplay between the translocation distances and transition rates for the folded and linear segments of the polymer. Theoretical results are applied for analysis of experimental translocations through solid-state nanopores.  
\end{abstract}

\maketitle

\end{titlepage}

\section{Introduction}

The motion of polymer molecules along channels is essential for many physical, chemical, biological and industrial processes, such as DNA and RNA transport across nuclear pores, viral infection, gene therapy and protein translocation via cellular membranes. \cite{lodish_book,meller_review} Recent experimental advances, that allowed to investigate the translocation of polymers with a single-molecule precision,\cite{kasianowicz96,henrickson00,meller00,meller01,meller02,sauer-budge03,bates03,mathe04,wang04,storm05,storm_storm,heng05,mathe05,mathe06,butler06,keyser06,murphy07,dekker_review} stimulated multiple theoretical and numerical studies of transport across the nanopores.\cite{sung96,lubensky99,muthukumar99,muthukumar02,ambjornsson02,slonkina03,flomenbom03,metzler03,kafri04,lakatos05,bundschuh05,chuang01,chern01,zandi03,milchev04,rabin05,muthukumar06,matysiak06,kotsev06,wolterink06,dubbeldam07} However, our understanding of fundamental processes underlying the translocation phenomena is still quite limited.

Most of experimental studies of polymer translocation are performed using $\alpha$-hemolysin pores embedded in  lipid bilayer membranes.\cite{kasianowicz96,henrickson00,meller00,meller01,meller02,sauer-budge03,bates03,mathe04,mathe05,mathe06,butler06,murphy07} Although protein channels provide a convenient method of investigation of the polymer transport, there are several limitations in application of this approach. Because the protein channels are slowly diffusing along  membranes, the time of experimental measurments is typically limited. In addition, $\alpha$-hemolysin are stable only for specific ranges of voltages, temperatures, PH and concentrations of different chemical compounds. To overcome these problems, artificial solid-state nanopores with controlled pore size and stability for wide range of external conditions have been developed and successfully applied for studying polymer translocation.\cite{storm05,storm_storm,heng05,keyser06,dekker_review}

In a typical translocation experiment the polymer is driven through the nanopore by external electric field that helps it to overcome the entropic barrier.\cite{meller_review} Because the diameter of narrow part of $\alpha$-hemolysin is less than 2 nm, only single-stranded DNA and RNA molecules can thread through these protein channels. At large voltages it was shown that the translocation time $\tau$ is proportional to the size of the polymer $N$, i.e., $\tau \simeq N$.\cite{meller_review} The experimental observations for the systems with  synthetic nanopores are rather different.\cite{storm05,storm_storm,heng05,keyser06,dekker_review} The most reliable data  currently can be obtained for solid-state pores with diameters ranging from 5 to 15 nm, that allow translocation of double-stranded DNA molecules, although smaller diameter channels have also been utilized.\cite{heng05} Because the diameter of solid-state pores is relatively large, it was observed recently\cite{storm05} that the polymer can move through the pore not only in the linear fashion but also in different folded states. In addition, the translocation time at large external fields scales as $\tau \simeq N^{\alpha}$ with $\alpha=1.27 \pm 0.03$.\cite{storm05,storm_storm} It was suggested that this behavior can be explained by hydrodynamic drag on the segments of the polymer outside of the pore.\cite{storm_storm} Although the motion of folded polymers through the nanopores has been observed experimentally,\cite{storm05} the theoretical description of this phenomenon is not available.

In this paper we investigate the translocation dynamics of polymers with folds using simple discrete stochastic models. We  analyze how the presence of folding configurations affects the overall translocation dynamics and scaling properties. Our theoretical method is an extension of the approach developed recently for analyzing the translocation of inhomogeneous polymers.\cite{kotsev06} The paper is organized as follows. The details of the model are given in Sec. II, the results of theoretical calculations are presented and discussed in Sec. III, and Sec. IV provides a summary and conclusions for our theoretical approach.

\section{Our Model}

We consider a motion of the single polymer molecule with $N$ monomers across the membrane pore as shown in Fig. 1. It is assumed that the polymer can translocate as a linear chain or in the configuration with a single fold, and during the translocation the length of the folded segment is constant. Experimental results indicate that the  transport of polymers with multiple folds (more than one) is negligible, at least for currently used solid-state nanopores,\cite{storm05} and we neglect this possibility. Each monomer has an effective charge $q$, and the polymer is driven through the nanopore by external electrostatic field $V$. To describe the part of the polymer in the folded state  we introduce a parameter $A$ ($0 \le A \le 0.5$) that specifies that the number of monomers in the folded segment to be 2$AN$: see Fig. 1. When $A=0$ the linear unfolded chain moves across the pore, while for $A=0.5$ the whole polymer molecule is in the folded state (folded at the middle of the chain). We also assume that the length of the nanopore is small so that only one (for the linear chains) or two (for the folded segments) monomers can be found inside the channel. This approximation is valid for the polymers with contour length much larger than the nanopore length. However, it was shown earlier that the effect of the nanopore length is important for translocation processes,\cite{meller_review,meller01,slonkina03} and it can be easily incorporated in our theoretical approach. In experiments\cite{meller_review,dekker_review} the external voltages are typically large, then as soon as the polymer enters the pore it has a very low probability of escaping back. That is why in this paper we neglect the exit back processes for translocating polymers.\cite{kotsev06}

When the polymer with folds moves across the membrane, first the folded segment translocates and then the linear part goes through the nanopore (see Fig. 1). Let us define $P_{k}(t)$ as a probability to have $k$ translocation events at time $t$. It is important to note that the parameter $k$ is generally different from the number of translocated monomers.  When the folded segment moves through the pore, at each translocation event two monomers are moved across the membrane, forward or backward. For linear segment during each event only one monomer is moved through the channel. Then the dynamics of the system is described by a master equation,
\begin{equation}\label{master}
\frac{dP_{k}(t)}{dt}=u_{k-1}P_{k-1}(t)+w_{k+1}P_{k+1}(t)-(u_{k}+w_{k}) P_{k}(t),
\end{equation}
where parameters $u_{k}$  and $w_{k}$ specify the transition rates to move the polymer molecule by one translocation event in the forward or backward direction, respectively.  These rates are related by  detailed balance conditions,
\begin{equation}\label{detailed_balance}
\frac{u_{k}} {w_{k+1}}=\exp[-\beta(F_{k+1}-F_{k})]=\exp[-\beta \Delta F_{k}],
\end{equation}
with $F_{k}$ corresponding to a free energy of the polymer  after $k$ successful forward translocation events (the number of translocated monomers can be larger), and $\beta=1/k_{B}T$.

The free energy of the polymer during the translocation process can be written as a sum of two terms, entropic and electrostatic,
\begin{equation}
F_{k}=F_{k,entr} +F_{k,elec}.
\end{equation}
However, theoretical calculations\cite{muthukumar99,slonkina03,kotsev06} show that the entropic contribution is much weaker than the electrostatic free energy at large external electric fields used in experiments,\cite{meller01,storm05} and it will be neglected in our theoretical calculations. The folded segments of the polymer molecule contribute even less into the entropic free energy, supporting the validity of this approximation. Because the translocating polymer consists of two blocks, folded and linear, the free energy will be different depending on what segment is currently in the pore,
\begin{equation}\label{free_energy1}
F_{k}=\left\{ \begin{array}{cc}
              -2kqV, & \mbox{ for } k <NA; \\
              -kqV -qVAN, & \mbox{ for } k >NA.
              \end{array} \right.
\end{equation}

In order to calculate the dynamic properties of translocating polymers we should have explicit expressions for the transition rates $u_{k}$ and $w_{k}$. However, the detailed balance conditions  (\ref{detailed_balance}) provide us only the ratio of these rates. By introducing a parameter $\theta$ ($0 \le \theta \le 1$), that specifies the distribution of free energy difference between the forward and backward transitions, the rates can be written in the following form,
\begin{equation}\label{wk}
u_{k}= D_{k} \exp[-\beta \theta \Delta F_k ], \quad w_{k+1}=  D_{k+1} \exp[\beta (1-\theta) \Delta F_k)],
\end{equation}
where $D_{k}$ corresponds to transition rates when $\Delta F_{k} =0$. Because the polymer molecule interacts stronger with the nanopore when the folded segment translocates through the membrane we assume that 
\begin{equation}
D_{k}=\left\{ \begin{array}{cc}
              BD, & \mbox{ for } k <NA; \\
              D, & \mbox{ for } k >NA.
              \end{array} \right.
\end{equation}
The parameter $B$ measures relative interaction strength  with the pore for the folded and linear segments. One can expect that the narrower the diameter of the nanopore, the stronger deviation of $B$ from unity. Then, utilizing the expressions for the free energy (\ref{free_energy1}), the translocation rates for the folded segment are
\begin{equation}\label{wkfold}
u_{k}=B\exp[\beta \theta 2qV], \quad w_{k+1}=B\exp[\beta (\theta -1) 2qV],  \mbox{ for } k <NA,
\end{equation}
while for the linear part the corresponding equations are given by
\begin{equation}\label{wkstraight}
u_{k}=\exp[\beta \theta qV], \quad w_{k+1}=\exp[\beta (\theta -1) qV],  \mbox{ for } k >NA,
\end{equation}

In experiments the translocation of the polymer is associated with current blockages of other ions in the system, and it is described by translocation times $\tau$. In our theoretical approach these times are associated with  mean first-passage times to cross the channel,\cite{kotsev06} and they can be calculated explicitly for different sets of parameters. The mean first-passage time  generally depends on the parameters $A$, $B$, the external voltage $V$ and the size of the polymer $N$. For the discrete stochastic model described by Eq. (\ref{master}), the standard expression for the mean first-passage time yields \cite{vanKampen,pury03} 
\begin{equation}\label{t_discret}
\tau=\sum_{k=1}^{M} \frac{1}{u_k} +  \sum_{k=1}^{M -1}\frac{1}{u_k}\sum_{i=k+1}^{M} \prod_{j=k+1}^{i} \frac{w_j}{u_j},
\end{equation}
where $M = N(1-A)$ is the translocation distance for the polymer with folded configurations, and it gives the maximal value for the parameter $k$.

\section{Results and Discussion}

The explicit expression for the translocation times through the nanopore can be obtained  by substituting  Eqs. (\ref{wkstraight}) and (\ref{wkfold}) into the general expression (\ref{t_discret}),
\begin{eqnarray}\label{polymer_discrete}
\tau&=&  \frac {NA \exp (-2\theta \beta qV)}{ B \left(1-\exp(-2\beta qV)\right)} +\frac{N(1-2A)\exp(-\theta\beta qV)}{\left(1-\exp(-\beta qV)\right)}\nonumber \\
&+&\frac {\exp \left(-(1+\theta )\beta qV\right)\left[\exp \left( -N(1-2A) \beta qV \right)-1 \right]}{\left( 1-\exp(-\beta qV) \right)^2} \nonumber \\
&+&\frac{\exp \left(-(2\theta + 1)\beta qV\right) \left[1+\exp\left(-\beta qVN \right)-\exp\left(-2\beta qVNA \right)- \exp\left(-(1-2A)\beta qVN \right)\right] }{ B \left(1-\exp(-\beta qV)\right)\left(1-\exp(-2\beta qV)\right)} \nonumber \\
&+&\frac{\exp \left(-2(1+\theta )\beta qV\right) \left[\exp (-2NA\beta qV) - 1\right]}{B\left(1-\exp(-2 \beta qV) \right)^2}.
\end{eqnarray}
The translocation time of the linear polymer without folds ($A=0$) is given by a simpler expression,
\begin{equation}\label{straight_discrete}
\tau_{0} = \frac{N \exp(-\theta \beta qV)}{1-\exp(-\beta qV) }+\frac{\exp \left(-(1+\theta) \beta qV \right)\left[\exp (-N\beta qV)-1 \right]} {\left(1-\exp(-\beta qV) \right)^2},\end{equation}
while the passage time for the fully folded polymer ($A=0.5$) can be written as
\begin{equation}\label{folded}
\tau_{f} = \frac {0.5N \exp (-2\theta \beta qV) }{ B\left(1-\exp(-2\beta qV)\right)} + \frac {\exp \left(-2(1+\theta)\beta qV\right) \left[\exp(-N\beta qV)-1 \right] }{ B\left( 1-\exp(-2 \beta qV) \right)^2}.
\end{equation}
Since we consider the motion of uniformly charged polymers, then it can be assumed that any monomer has equal probability to be captured by the nanopore. Then the probability for the polymer to be found in one of $N$ possible  configurations is the same. Averaging out over this uniform distribution, i.e., integrating  over the variable $0 \le A \le 0.5$, leads us to the mean passage time through the nanopore for the given polymer size $N$, 
\begin{eqnarray}\label{average_discrete}
\langle \tau \rangle &=&\frac {N \exp(-2 \theta \beta qV)} {4B (1-\exp(-2\beta qV))}+  \frac{N \exp\left(-\theta \beta qV \right)}{2(1-\exp(-\beta qV))} \nonumber \\
&+&\frac {\exp \left(-(1+\theta  )\beta qV\right) \left[1- \exp (- \beta qVN) - \beta qVN \right]}{\beta qVN \left(1-\exp(-\beta qV) \right)^2}   \nonumber \\
&+&\frac {\exp \left(-2(1+\theta)\beta qV\right)\left[1- \exp (- \beta qVN) - \beta qVN \right] }{B \beta qVN \left( 1-\exp(-2\beta qV) \right)^2} \nonumber \\
&+&\frac {\exp \left(-(1+2\theta)\beta qV\right) \left[1+\exp(-\beta qVN) +\frac{2}{\beta qVN}\left(\exp\left(-\beta qVN \right)-1\right)\right]}{B\left(1-\exp(-\beta qV) \right)\left( 1-\exp(-2\beta qV) \right)}. 
\end{eqnarray}
Experimental measurements of the capture position along the polymer chain  in solid-state nanopore translocations\cite{storm05} support the assumption of the uniform distribution of fold locations. However, at experimental conditions linear (unfolded) configurations appear 10 times more frequently, and for large DNA the fully folded configuration ($A=0.5$) is also twice more probable. This last observation can be explained by a tendency of DNA to make circular molecules.\cite{storm05} Our theoretical approach can easily takes into account these more realistic distributions of the folding configuration in the polymer translocation. 

The results of theoretical calculations for the relative translocation times $\tau/\tau_{0}$ as a function of the length of the folded segment $A$ are shown in Fig. 2. When the motion of the folded segment is significantly hindered due to large interactions with the nanopore ($B=0.01$ in Fig. 2a), the polymers with  folded states generally spend more time in the pore than the unfolded molecules. It is interesting to note a non-monotonous behavior of translocation times for relative small polymer sizes and weak external fields ($N=10, \beta qV=0.1$ in Fig. 2a). This can be explained by considering the details of the translocation process for the folded and linear configurations. The folded segment of the molecule moves slower than the linear part, however the translocation distance $M$ decreases with increasing $A$, $M=N(1-A)$. Thus for $A$ close to 0.5 the polymer with folded configurations lowers its translocation time. The increase in the polymer size $N$ and/or the increase in the external voltage washes out this effect (see Fig. 2a). For  large voltages the folded polymer can even spend less time in the channel. If the polymer interacts weakly with the nanopore ($B=1$ in Fig. 2b) the folded polymers always  translocates faster than the linear chains because of the decrease in the translocation distance.  

The effect of interactions between the polymer and the nanopore on translocation dynamics is shown in Fig. 3. The increase in the parameter $B$, that corresponds to the lowering the interaction with the channel, accelerates the dynamics of the polymers with folded configurations, as expected. The results in Fig. 3 indicate that for any fixed value of the parameter $A$ there is a critical value of interaction  strength $B^{*}$ that describes two different translocation dynamic behaviors. For $B<B^{*}$ the linear chain is the fastest polymer configuration, while for $B>B^{*}$ the folded polymer translocates faster. The value of the parameter $B^{*}$ for given $A$ can be found from the equation $\tau(A)=\tau_{0}$. Specifically, for the fully folded configurations ($A=0.5$) the critical parameter can be derived from Eqs. (\ref{straight_discrete}) and (\ref{folded}), yielding
\begin{equation}\label{criticalB}
B^{*}=\frac{\frac {0.5N \exp (-2\theta \beta qV) }{\left(1-\exp(-2\beta qV)\right)} + \frac {\exp \left(-2(1+\theta)\beta qV\right) \left[\exp(-N\beta qV)-1 \right] }{\left( 1-\exp(-2 \beta qV) \right)^2}}{\frac{N \exp(-\theta \beta qV)}{1-\exp(-\beta qV) }+\frac{\exp \left(-(1+\theta) \beta qV \right)\left[\exp (-N\beta qV)-1 \right]} {\left(1-\exp(-\beta qV) \right)^2}}.
\end{equation}

The value of the critical interaction depends on the size of the polymers and external voltages, as illustrated in Fig. 4. In the limit of $N \gg 1$ the dependence of $B^{*}$ on the polymer's size disappears, and the expression (\ref{criticalB}) simplifies into
\begin{equation}
B^{*} \simeq \frac{0.5 \exp(-\theta \beta qV)}{1+ \exp(-\beta qV)}.
\end{equation}
Then, in this limit, for $\theta=0.5$ we  calculate that $B^{*} \simeq 0.22$ for $\beta qV=1$, and $B^{*} \simeq =0.003$ and $B^{*} \simeq 0.25$ for $\beta qV=10$ and 0.1, respectively. As shown in Fig. 4a, the value for the critical interaction already saturates for $N>40$. The critical interaction is also independent of the polymer size in the limit of large external fields (see Fig. 4b), when $B^{*} \simeq 0.5 \exp(-\theta \beta qV)$. The size dependence effectively disappears for $\beta qV >2$ (for $\theta=0.5$).

Our theoretical predictions can be compared with experimental observations of translocation dynamics  for DNA molecules of sizes 11.5 kbp and 48.5 kbp.\cite{storm05} In this experiments the translocation times for linear and fully folded polymers have been measured. Since the sizes of DNA molecules are large, we can use Eqs.  (\ref{straight_discrete}) and (\ref{folded}) for determination of the relative interaction strength between DNA and the solid-state nanopore,
\begin{equation}\label{Bexp}
B_{exp}=\frac{(\tau_{0}/\tau_{f})\exp(-\theta \beta qV)}{1+ \exp(-\beta qV)}.
\end{equation} 
The experiments indicate that $ \tau_{0}/\tau_{f}$ is approximately equal to $2.04 \pm 0.12$ for the DNA  with the length of 11.5 kbp, while  $ \tau_{0}/\tau_{f} \approx 2. 18 \pm 0.10$ for the DNA molecule with the length of 48.5 kbp.\cite{storm05} The fact that this ratio is almost the same for both  polymers is in agreement with our theoretical predictions because both molecular sizes are large. The experiments have been performed at $V=120$ mv. However, it is difficult to evaluate  the effective charge $q$ because of possible electrostatic screening and condensation effects.\cite{rabin05,keyser06} We estimate that $1 \le \beta qV \le 10$.  Then from Eq. (\ref{Bexp}) one can calculate the value of the parameter $B$ that characterizes the interaction between the DNA molecule and the nanopore. Our calculations (for $\theta=0.5$) yield the estimate of  $0.01 < B < 0.20$, suggesting that there is an interaction that slows the translocation of the folded part of the molecule. However this interaction is weak enough so that the folded polymers translocation time is similar to the linear chain at this external voltage, in agreement with experimental observations.\cite{storm05} Eq. (\ref{Bexp}) also implies that by measuring the effective interactions for different voltages it is  possible, in principle, to estimate the value of the effective charge. 

Experiments on polymer translocation can measure the scaling properties of translocating polymers. It was found\cite{meller_review,meller01,storm05} that the passage time has a power-law dependence, $\tau \propto N^{\alpha}$, with an exponent $\alpha=1$ for the translocation across $\alpha$-hemolysin, and this observation agreed with some theoretical predictions.\cite{muthukumar99,slonkina03} However, surprisingly, for the transport across the solid-state nanopores the exponent was found to be different, $\alpha=1.27$.\cite{storm05} It was argued theoretically that the hydrodynamic drag on the sections of the polymer outside of the nanopore can explain this result.\cite{storm_storm} One could also suggest that the presence of the folded configurations might  change the polymer scaling properties.  Our theoretical method allows us to test this possibility. A dependence of the relative mean translocation time (averaged over all folding configurations) as a function of the polymer size $N$ for different interactions between the polymer and the nanopore is presented in Fig. 5. Because the function $\langle \tau \rangle /\tau_{0}$ is independent of the polymer size at large $N$, we conclude that folded configurations do not affect the scaling properties.

\section{Summary and Conclusions}

We developed a simple discrete stochastic model to describe the translocation of polymers with folded configurations. This approach allowed us to obtain exact analytical expressions for translocation times as functions of the length of the folded segments,  interaction between the polymer and the nanopore, the polymer size and the external electric field. It is shown that for large interactions that significantly slow  down the motion of the folded segments, the linear chains move faster, although the dependence on the folded fraction is non-monotonous for not very large polymers and external fields. If the interaction is weak, the folded polymers always translocate faster than the linear unfolded polymers, and this effect is even stronger for large external voltages. Our theoretical analysis predicts that there is a critical interaction that separates two translocation regimes, depending on which  linear or folded configurations are the fastest. This critical interaction generally depends on the length of the polymer molecule and the external voltage, although at large $N$ and/or large $V$ the critical parameter becomes independent of the polymer size. Also, theoretical calculations support the arguments that the existence of folded configurations does not change the scaling properties of the polymers. These theoretical observations are explained by analyzing the translocation distances and speeds for different polymer configurations. Our theoretical results are utilized for the description of experiments on translocation of polymers with folded configurations.

Presented theoretical description of the translocation of polymers with folded configurations is based on the oversimplified model. To develop a more realistic approach to this complex problem several important properties should be accounted for. First of all, the entropic free energy contributions must be included. However, it is unlikely that qualitative predictions obtained in this work would change in this case because for the realistic systems the entropic contributions are relatively weak.\cite{muthukumar99,slonkina03,kotsev06} Much more serious is the fact that the translocation in this work  is viewed as Markov process without memory  that can be described by ordinary master equations. Recent theoretical and computation studies of polymer translocation\cite{wolterink06,dubbeldam07} suggest that memory effects play important role in translocation, at least for low external force regimes. It is possible that this non-Markovian behavior can be taken into account by considering continuous-time random walks approach with generalized master equations.\cite{kolomeisky00} However, this problem deserves a more careful investigation. Also, our theoretical method does not take into account the sequence dependence and hydrodynamic effects. It will be interesting to investigate further the translocation of polymer molecules with folded segments by experimental and theoretical methods.

\section*{Acknowledgments}

The authors would like to acknowledge the support from the Welch Foundation (Grant No. C-1559), the U.S. National Science Foundation (Grants No. CHE-0237105 and ECCS-0708765). A.B.K. is also grateful to C. Clementi, M. Pasquali and D. Panja for valuable discussions.

\newpage

\noindent {\bf Figure Captions:} \\
\vspace{5mm}

\noindent Fig. 1. Schematic view of translocation of a polymer with a folded configuration. The size of the polymer molecule is $N$, and 2$AN$ is the number of monomers in the folded segment.

\vspace{5mm}

\noindent Fig. 2. Relative translocation time as a function of the length of the folded segment for different polymer sizes and different external fields: a) $B=0.01$; and b) $B=1$. For all calculations $\theta=0.5$ is assumed.

\vspace{5mm}

\noindent Fig. 3. Relative translocation time as a function of interaction strength between the polymer and the nanopore  for different polymer sizes and different external fields. For all calculations $\theta=0.5$ is assumed.

\vspace{5mm}

\noindent Fig. 4. Critical interaction parameter $B^{*}$ as a function of a) polymer size and b) external voltage. For all calculations $\theta=0.5$ is assumed.

\vspace{5mm}

\noindent Fig. 5. Relative mean passage time for the polymer with folded configurations as a function of the polymer size: a) $B=0.1$; and b) $B=10$. For all calculations $\theta=0.5$ is assumed.

\newpage

\begin{figure}[ht]
\begin{center}
\unitlength 1in
\begin{picture}(3.0,4.0)
  \resizebox{3.375in}{3.375in}{\includegraphics{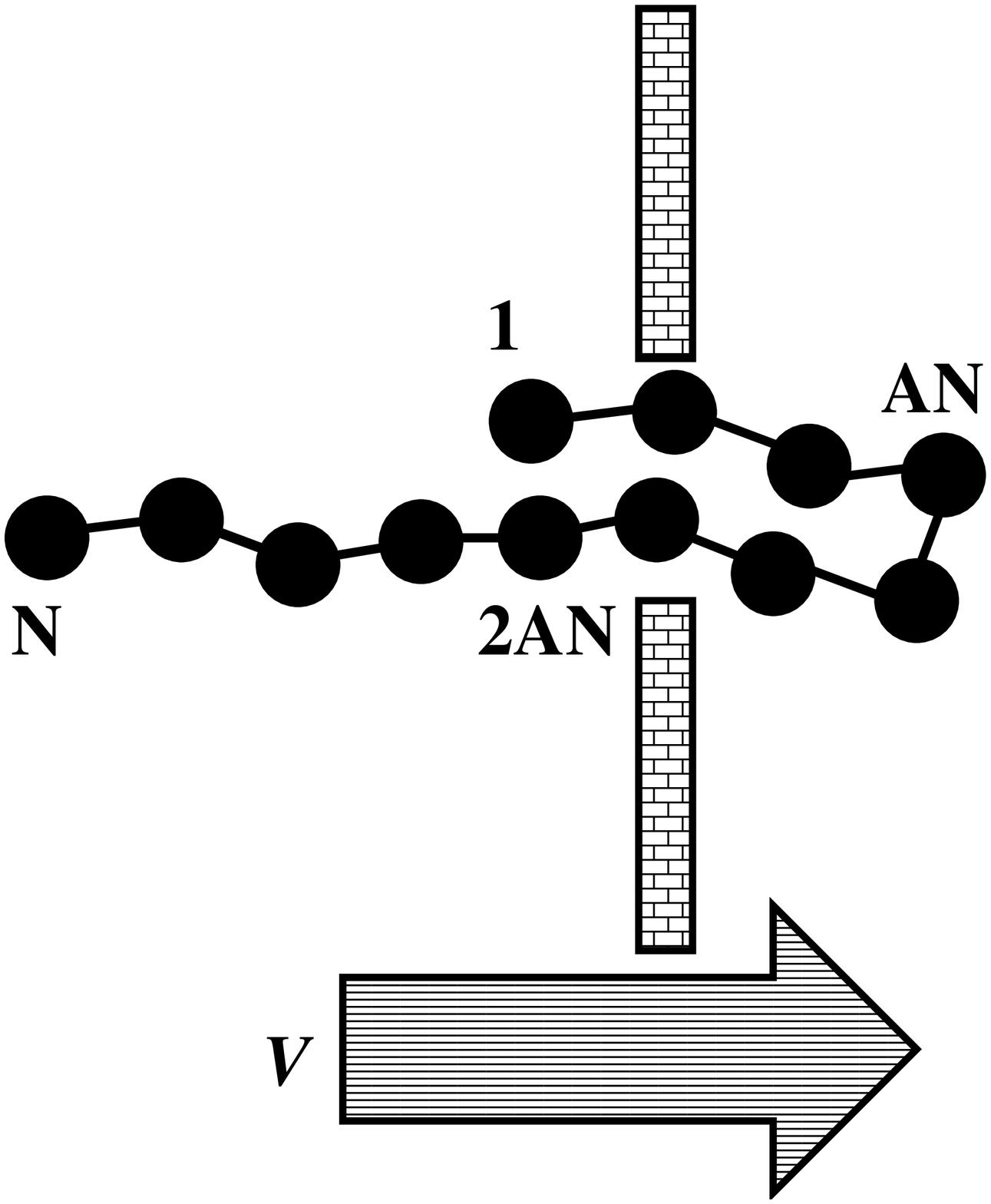}}
\end{picture}
\vskip 1in
 \begin{Large} Figure 1. Kotsev and Kolomeisky \end{Large}
\end{center}
\end{figure}

\newpage

\begin{figure}[ht]
\begin{center}
\unitlength 1in
\begin{picture}(3.0,4.0)
  \resizebox{3.375in}{3.375in}{\includegraphics{Fig2a.fold.eps}}
\end{picture}
\vskip 1in
 \begin{Large} Figure 2a. Kotsev and Kolomeisky \end{Large}
\end{center}
\end{figure}

\newpage

\begin{figure}[ht]
\begin{center}
\unitlength 1in
\begin{picture}(3.0,4.0)
  \resizebox{3.375in}{3.375in}{\includegraphics{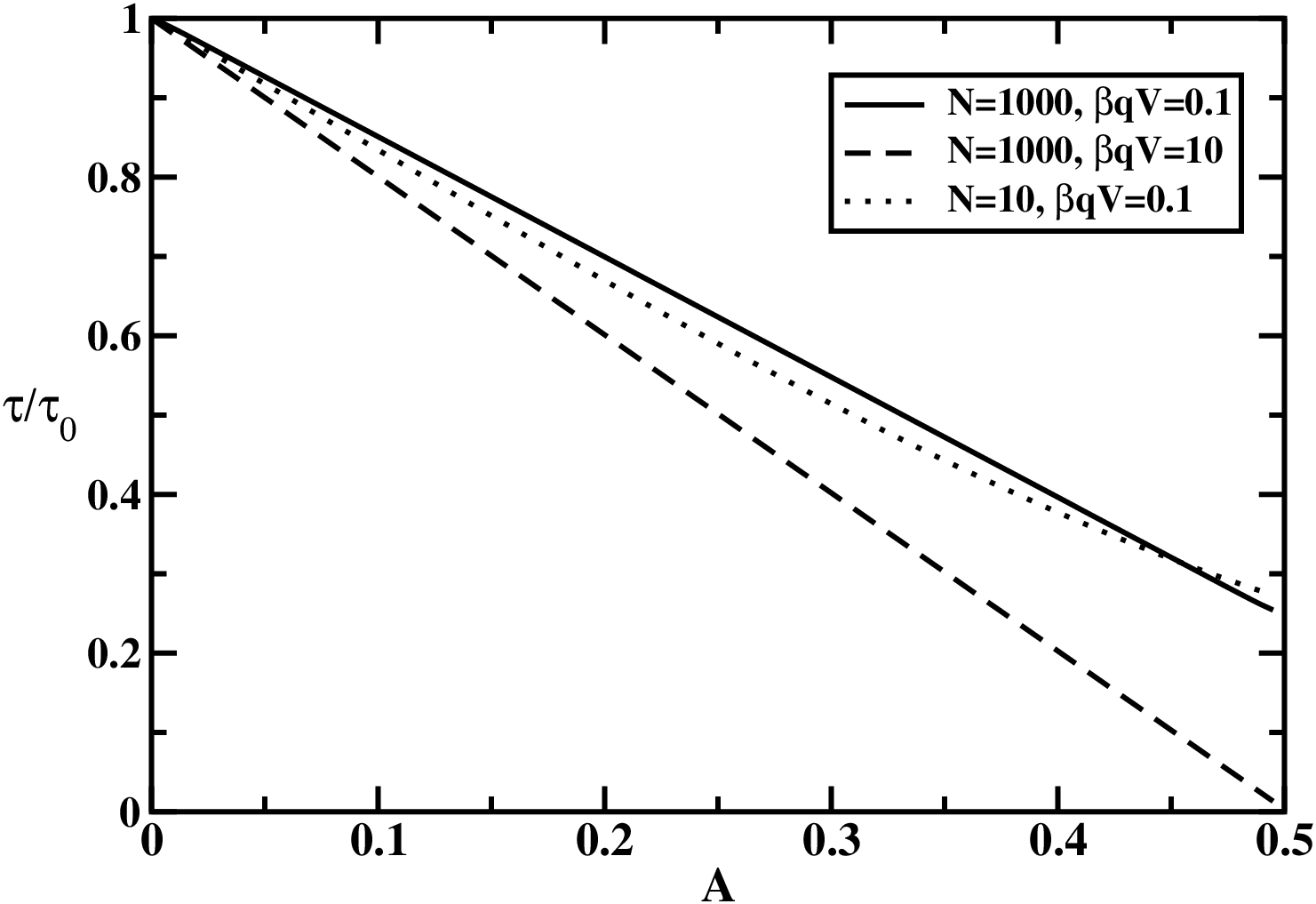}}
\end{picture}
\vskip 1in
 \begin{Large} Figure 2b. Kotsev and Kolomeisky \end{Large}
\end{center}
\end{figure}

\newpage

\begin{figure}[ht]
\begin{center}
\unitlength 1in
\begin{picture}(3.0,4.0)
  \resizebox{3.375in}{3.375in}{\includegraphics{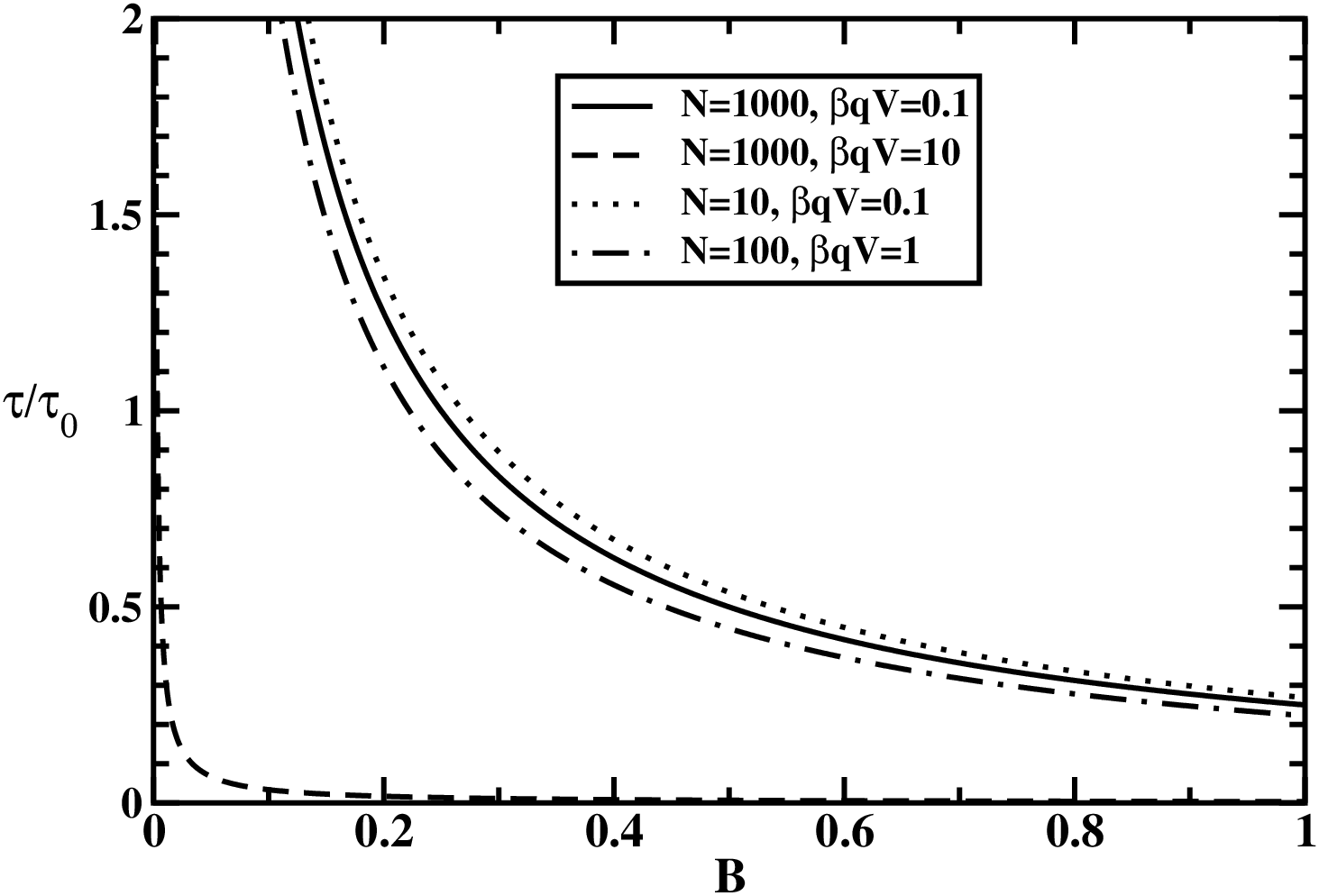}}
\end{picture}
\vskip 1in
 \begin{Large} Figure 3. Kotsev and Kolomeisky \end{Large}
\end{center}
\end{figure}

\newpage

\begin{figure}[ht]
\begin{center}
\unitlength 1in
\begin{picture}(3.0,4.0)
  \resizebox{3.375in}{3.375in}{\includegraphics{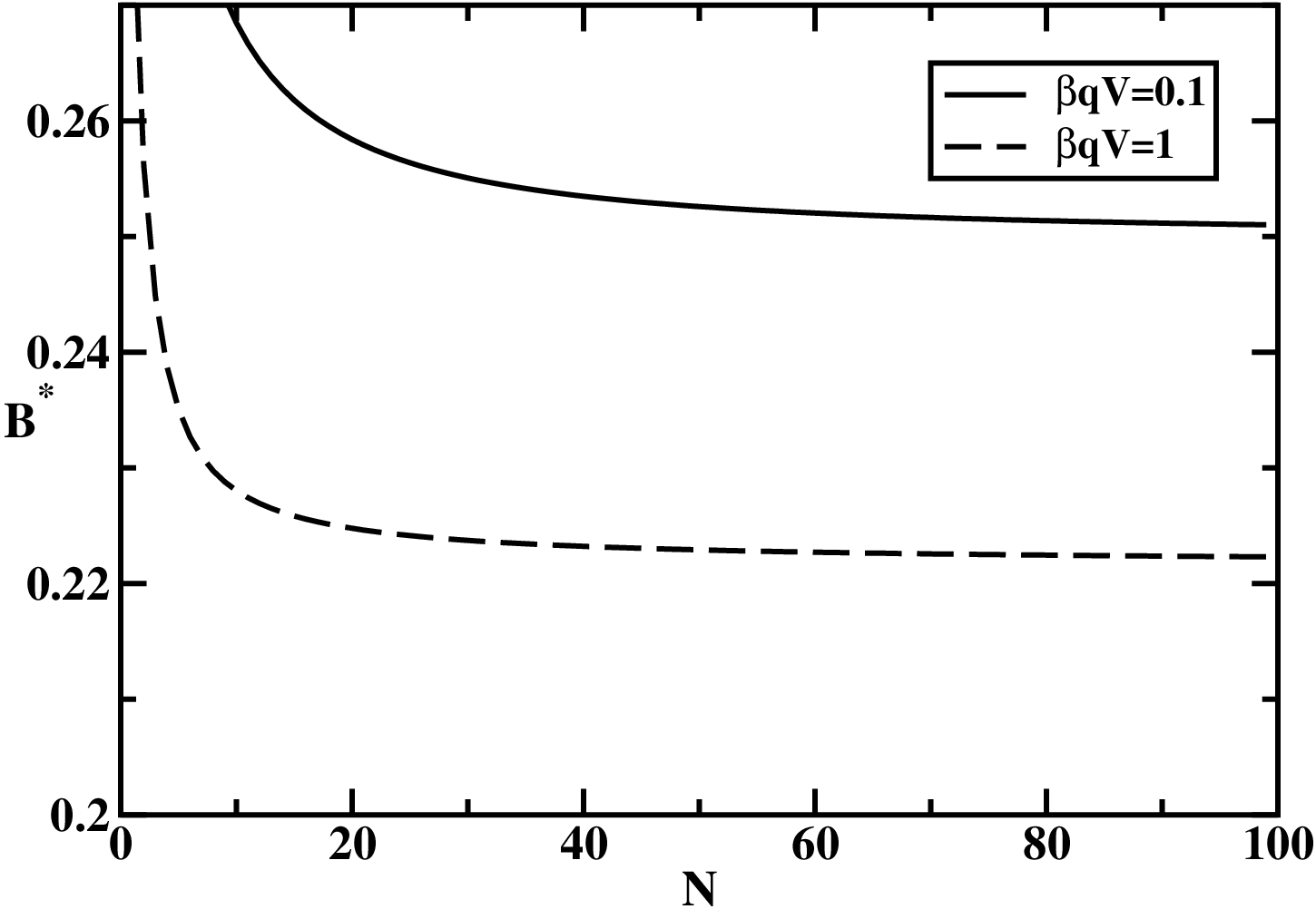}}
\end{picture}
\vskip 1in
 \begin{Large} Figure 4a. Kotsev and Kolomeisky \end{Large}
\end{center}
\end{figure}

\newpage

\begin{figure}[ht]
\begin{center}
\unitlength 1in
\begin{picture}(3.0,4.0)
  \resizebox{3.375in}{3.375in}{\includegraphics{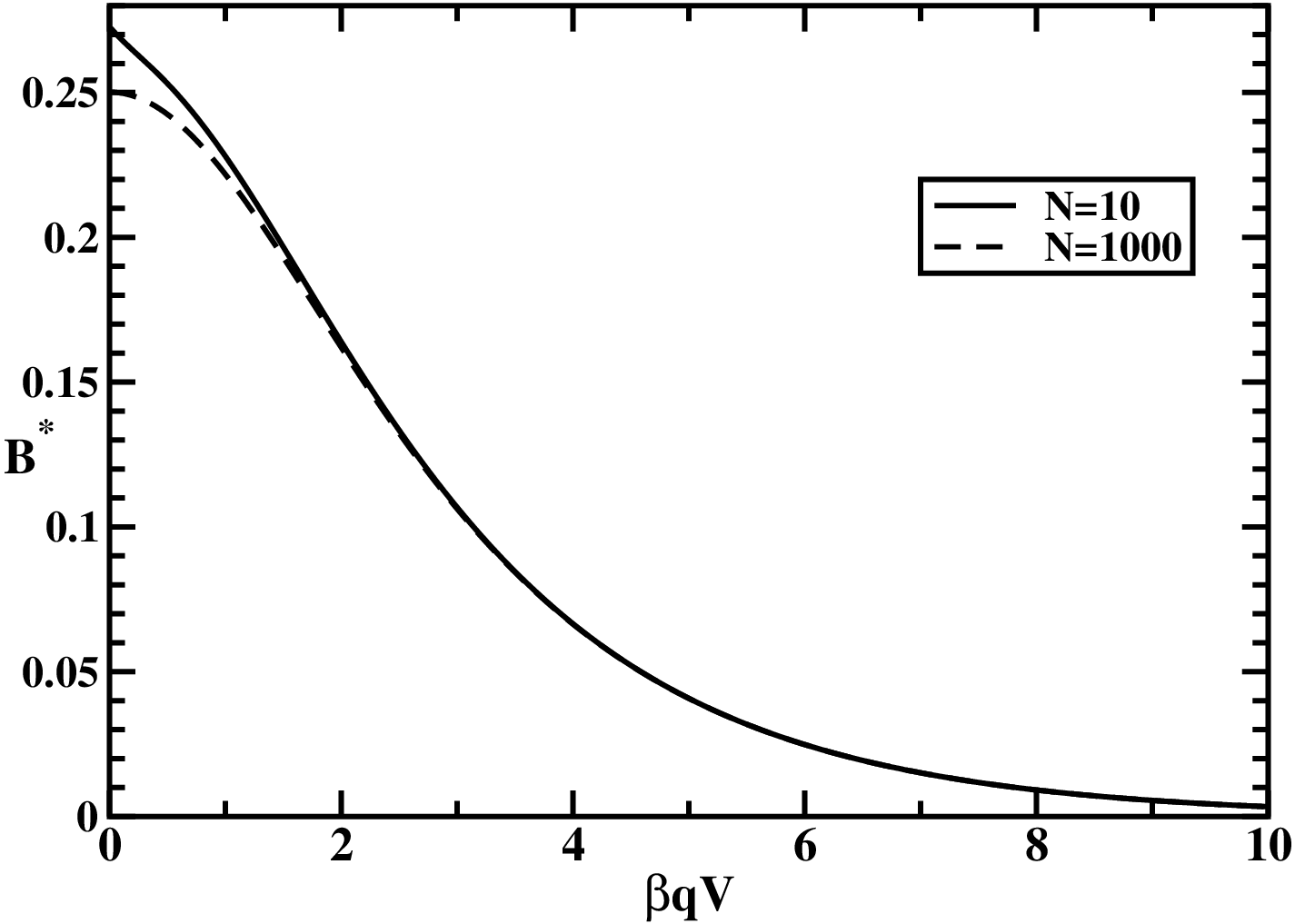}}
\end{picture}
\vskip 1in
 \begin{Large} Figure 4b. Kotsev and Kolomeisky \end{Large}
\end{center}
\end{figure}

\newpage

\begin{figure}[ht]
\begin{center}
\unitlength 1in
\begin{picture}(3.0,4.0)
  \resizebox{3.375in}{3.375in}{\includegraphics{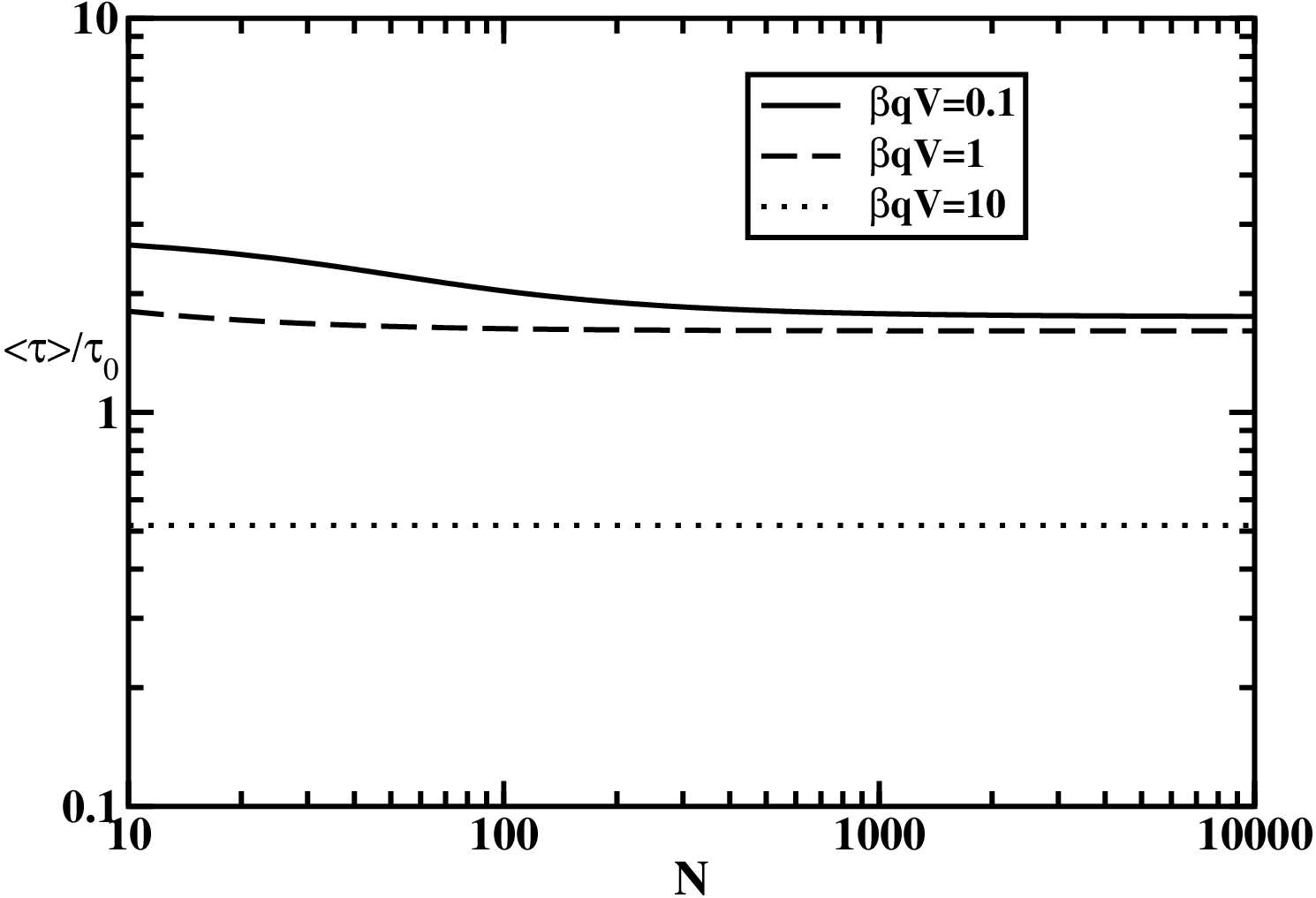}}
\end{picture}
\vskip 1in
 \begin{Large} Figure 5a. Kotsev and Kolomeisky \end{Large}
\end{center}
\end{figure}

\newpage

\begin{figure}[ht]
\begin{center}
\unitlength 1in
\begin{picture}(3.0,4.0)
  \resizebox{3.375in}{3.375in}{\includegraphics{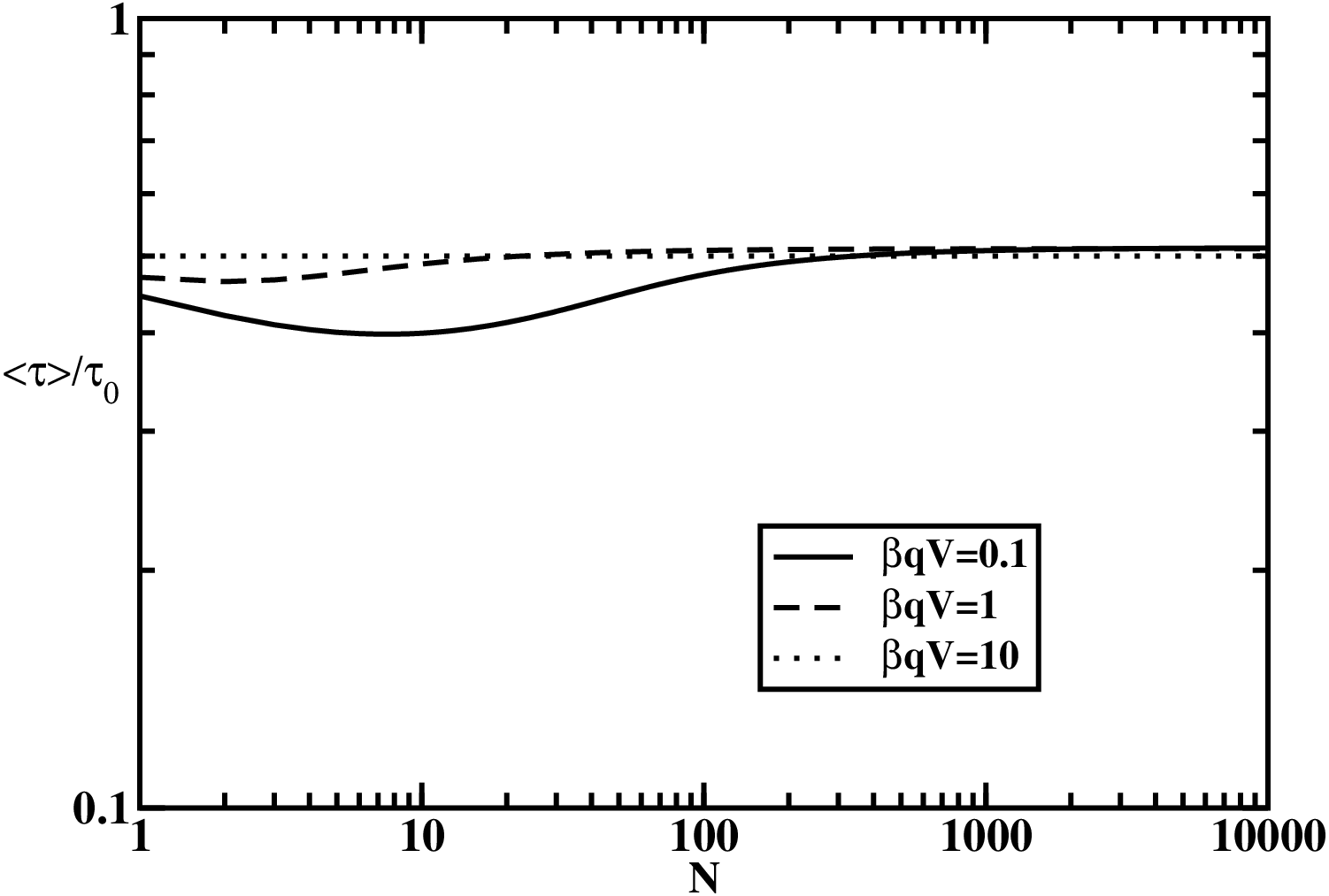}}
\end{picture}
\vskip 1in
 \begin{Large} Figure 5b. Kotsev and Kolomeisky \end{Large}
\end{center}
\end{figure}

\end{document}